\documentclass{caosp305}

\usepackage{graphicx}

\usepackage{natbib}
\articleNo{1}
\pubyear{2017}
\volume{00}
\volnumber{0}
\firstpage{1}
\received{October 31, 2017}
\accepted{October 31, 2017}

\def\BibTeX{{\rm B\kern-.05em{\sc i\kern-.025em b}\kern-.08em
             T\kern-.1667em\lower.7ex\hbox{E}\kern-.125emX}}

\begin{document}

\hauthor{A.\,Kawka}

\title{The properties and origin of magnetic fields in white dwarfs}

\author{A.\,Kawka \inst{1}}

\institute{\ondrejov, \email{kawka@asu.cas.cz}}

\date{October 17, 2017}

\maketitle

\begin{abstract}
A significant fraction of white dwarfs harbour a magnetic field with 
strengths ranging from a few kG up to about 1000 MG. The fraction appears to
depend on the specific class of white dwarfs being investigated and may hold
some clues to the origin of their magnetic field. 
The number of white dwarfs with variable fields as
a function of their rotation phase have revealed a large field structure
diversity, from a simple offset dipole to structures with spots or multipoles.
A review of the current challenges in modelling white dwarf atmospheres 
in the presence of a magnetic field is presented, and the
proposed scenarios for the formation of magnetic fields in white dwarfs are examined.
\end{abstract}

\section{Introduction}

\label{intr}

White dwarfs represent the final stage of stellar evolution for the majority
of stars and a significant fraction of them harbour a magnetic field ranging 
from a few kG up to several hundred MG. The presence of a magnetic field affects
the external appearance of the white dwarf (emerging flux), its temperature structure
as well as its evolutionary prospects (cooling age).

In this review I examine the incidence of magnetism in the white dwarf
population, and how it differs between different subclasses. The challenges
in modelling white dwarf atmospheres in the presence of a magnetic field
are discussed in Sect.~3. Finally, eligible scenarios for the origin of magnetic
fields in white dwarfs are explored in Sect. 4.

\section{Magnetic field incidence}

The measured fraction of magnetism in white dwarfs varies between various 
surveys. Magnetic white dwarfs are identified via polarization measurements (P)
or characteristic Zeeman patterns (H). Colourimetric and photometric limited surveys 
\citep[e.g.,][]{sch1995,kep2013} delivered fractions as low as 5\% but volume 
limited surveys \citep{kaw2007} resulted in fraction estimates as high as 20\%. 
The number of known magnetic white dwarfs has grown considerably in recent years
which helped uncover specific classes of white dwarfs showing a significantly higher
incidence of magnetism.

Most white dwarfs have a hydrogen-rich atmosphere (DA), with the remainder 
having an atmosphere dominated by helium, with a visual helium spectrum (DB) or
without (DC). A few, rare carbon-dominated objects are known as hot DQ white 
dwarfs \citep{duf2008} that appear to be high temperature counterparts to the 
cool helium-dominated but carbon-polluted DQ white dwarfs \citep{duf2005}. 
Approximately 25 to 30\% of white dwarfs show traces of heavy elements 
\citep{zuc2003,zuc2010} and for such objects the suffix Z is added to the
respective spectral classes, e.g., DAZ, DBZ, DZ (i.e., DC with metal lines).

\begin{figure}[!t]
\centerline{\includegraphics[width=0.75\textwidth,clip=20 500 595 540]{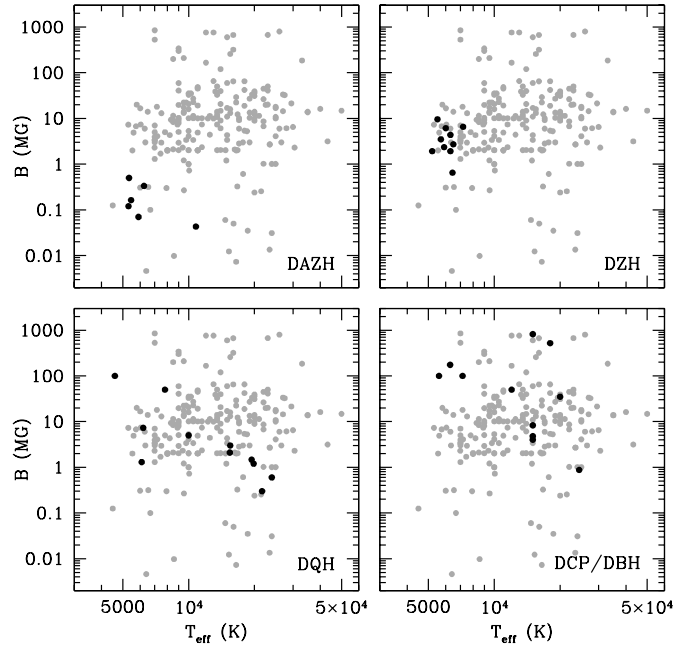}}
\caption{The distribution of known white dwarfs as a function of the magnetic 
field and effective temperature. The grey points in all panels are magnetic DA white 
dwarfs. Each panel highlights (in black) the properties of the various spectral type of 
magnetic white dwarfs (DAZ, DZ, DQ and DC/DB).}
\label{fig_mag_dist}
\end{figure}

Fig.~\ref{fig_mag_dist} shows all currently known magnetic white dwarfs 
segregated into various spectral types as a function of their field strength
and effective temperature. Table~\ref{tbl_class} lists the number of known 
magnetic white dwarfs and the incidence of magnetism per spectral class. 
For hot DQs, 12 are known, but only 6 of these have a published magnetic 
field measurement. There is no apparent correlation between effective 
temperature and magnetic field strength for DAH and DCP/DBH white dwarfs. 
Although only a few are known, polluted magnetic white dwarfs (DAZH, DZH) 
are clustered at cooler temperatures, with the DAZH at systematically lower 
field strengths. Featureless DC white dwarfs are difficult to diagnose and 
only a handful of high-field DCP are known.

\begin{table}[t!]
\caption{Incidence of magnetism among different classes of white dwarfs}
\label{tbl_class}
\vspace{-0.5cm}
\begin{center}
\footnotesize
\renewcommand{\arraystretch}{1.25}
\begin{tabular}{lcccc}
\hline\hline
Spectral type & Prototype & Number & Fraction (\%) & Reference \\
\hline
DAH        & Grw+70$^\circ$8247 & 208    & $4\pm1.5$ & \citet{sch1995} \\
DAZH       & G77-50             & 6      & $\sim 50$ & Table 2 \\
DBH        & GD229              & 8      & $\sim 1.5$& estimated \\
DCP        & G195-19            & 3      & $\sim 5$  & \citet{put1997} \\
DZH        & LHS2534            & 10     & $13\pm4$  & \citet{hol2015} \\
DQH (hot)  & SDSS~J1337+0026    & 12 (6) & $\sim 70$ & \citet{duf2013} \\
DQH (cool) & G99-37             & 4      & $\sim4$   & \citet{vor2013} \\
\hline\hline
\end{tabular}\\
\renewcommand{\arraystretch}{1}
\end{center}
\end{table}

\subsection{Cool polluted white dwarfs}

Studies of cool, polluted white dwarfs have revealed a higher incidence of
magnetism than in the general population of white dwarfs. \citet{kaw2014} 
showed an incidence of 40\% in cool, polluted hydrogen-rich (DAZ) white dwarfs.
These objects have relatively low fields ($B_S < 1$~MG).
Here we revisit this sample and confirm this high incidence. Table~\ref{tbl_daz}
lists the known magnetic DAZs. Comparing this sample to all the other known DAZ 
white dwarfs that have been observed at sufficiently high resolution we find
that close to 50\% of DAZ white dwarfs with $T_{\rm eff} < 6000$ K are magnetic. 
The abundance patterns of magnetic DAZs does not appear to differ from those
of non-magnetic DAZs and we cannot establish a correlation between magnetic field
strengths and abundances.
Above a temperature of 6000~K, only 2 magnetic DAZs are known, NLTT~53908 is only slightly
warmer at 6250~K and WD2105-820 is hotter with $T_{\rm eff} = 10\,800$~K. 

A high incidence of magnetism is also observed in the cool, polluted, 
helium-rich class (DZ) of white dwarfs. \citet{hol2015} reported an incidence 
of 13\% in DZ white dwarfs with $T_{\rm eff} < 8000$~K. The magnetic fields
in this case are higher than those in cool DAZs with $1.9 < B_S < 9.6$ MG. 
\citet{hol2015} noted that the incidence of magnetism in DZs hotter than 8000 K
is significantly lower in their sample.

\begin{table}[t]
\caption{Properties of known magnetic DAZ white dwarfs}
\label{tbl_daz}
\vspace{-0.3cm}
\begin{center}
\footnotesize
\renewcommand{\arraystretch}{1.25}
\begin{tabular}{lccccc}
\hline\hline
Name & $T_{\rm eff}$ (K) & $\log{g}$ & $B_S$ (kG) & V (mag) & Reference \\
\hline
NLTT07547 & 5460 & 8.04 & 163 & 18.3 & 1 \\
NLTT10480 & 5410 & 8.0  & 519 & 17.5 & 2 \\
NLTT43806 & 5900 & 8.0  & 70  & 15.9 & 3,4 \\
NLTT53908 & 6250 & 7.87 & 334 & 18.0 & 5 \\
G77-50    & 5310 & 8.05 & 120 & 16.2 & 6 \\
WD2105-820 & 10800 & 8.19 & 43 & 13.6 & 7,8 \\
\hline\hline
\end{tabular}\\
References: (1) Kawka et al., in prep; (2) \citet{kaw2011}; \\
(3) \citet{kaw2006}; (4) \citet{zuc2011}; (5) \citet{kaw2014}; \\ 
(6) \citet{far2011}; (7) \citet{koe2009}; (8) \citet{lan2012}
\renewcommand{\arraystretch}{1}
\end{center}
\end{table}

\subsection{Hot DQ white dwarfs}

Hot DQ white dwarfs have temperatures ranging from about 18\,000~K up to
24\,000~K and an atmosphere dominated by carbon. Half of these stars were 
found to be photometrically variable with periods ranging from $\approx 5$ min
up to 2.1 days. Initially, these variations were attributed to pulsations, however,
following the discovery of a 2.1 day period in SDSS~J0005-1002 \citep{law2013}, the 
preferred explanation is that they are caused by a rotating field. Recently, \citet{duf2013}
reported an incidence of 70\% in this class of objects, suggesting that
all hot DQs may be magnetic at some level. The magnetic fields range from
$\sim 0.3$ up to 2.1~MG. \citet{duf2013} also propose that
these stars may be more massive than the general white dwarf population.
\citet{dun2015} also suggests that hot DQs are likely massive, and, since most
of them show rapid variability which may be attributed to rotation,
they are possibly the product of white dwarf mergers.

\section{Modelling magnetic atmospheres}

Magnetic fields, particular strong ones, are detected in intensity spectra while
weak fields are readily detectable in Stokes $V$ spectra. A combination of 
intensity and circular polarization spectra are useful for detailed studies of
field geometry. Different Zeeman regimes need to be used depending on
the field strength. The linear Zeeman regime can be assumed for weak fields, and
as the field strength increases, the quadratic effect becomes important. For the
strongest magnetic fields the strong field mixing regime needs to be adopted. For 
more details about these regimes see \citet{wic2000}.

In general the magnetic field in white dwarfs is assumed to be a centred or offset
dipole. However, rotating magnetic white dwarfs have revealed a diversity in the
field topology. \citet{lan2017} showed that the rotating magnetic white dwarf 
WD~2047+372 which has a weak field of $B_P = 91.8\pm0.8$~kG
can be modelled by a simple dipole. Their analysis of a second low-field white dwarf
WD~2359-434 required the combination of a dipolar and a non-aligned quadrupole to 
model the spectropolarimetric observations. Some white dwarfs were found to have
even more complex structures. For example, the high-field, massive and hot 
white dwarf EUVE~J0317-855 shows a field of 185 MG with a likely 425 MG 
magnetic spot \citep{bur1999,ven2003}. WD~1953-011 also shows a complex field 
structure \citep{max2000,val2008} as it rotates with a period of 1.4418 days 
\citep{bri2005}.

White dwarfs with hydrogen-rich atmospheres become convective below 
$T_{\rm eff} \sim 15\,000$~K. Helium-rich atmospheres develop convection zones
at much higher temperatures $T_{\rm eff} \la 30\,000$~K.
\citet{val2014} proposed that convection is suppressed in magnetic white dwarfs
and slows down the white dwarf cooling rate.
Using radiative magnetohydrodynamic simulations \citet{tre2015} confirmed that
convection is indeed suppressed, but that the cooling rate is not affected until
the convective zone couples with the degenerate core which occurs around 5500~K.
Observationally, \citet{gen2017} fitted far ultraviolet (UV) spectra as well
as optical Balmer line spectra of magnetic and non-magnetic white dwarfs with
temperatures ranging from 9000 to 10\,000~K. They used models with convection
and without convection and found that for the magnetic white dwarf
WD~2105-820 they could only produce consistent results between the best fitting UV and 
optical data using radiative models. 

\section{Origin of magnetic fields in white dwarfs}

Magnetic fields in white dwarfs have often been assumed to be fossil fields,
and that the progenitors of magnetic white dwarfs are predominantly magnetic Ap and Bp stars.
Assuming magnetic flux conservation, the 
field strengths observed in Ap/Bp stars would correspond to white dwarf 
fields in excess of 10 MG \citep{kaw2004,wic2005}.
Therefore, the progenitors of white dwarfs with weak magnetic fields may
be other main-sequence stars that have magnetic fields well below current 
detection limits. 

The fossil field theory implies that magnetic white dwarfs should be found in
equal proportion in binary systems with main-sequence stars. However, this 
does not appear to be the case. Extensive surveys, like those of the Sloan 
Digital Sky Survey (SDSS), have been unable to find any non-interacting 
magnetic white dwarf plus main-sequence pairs \citep{lie2015}, thus leaving 
magnetic cataclysmic variables without direct progenitors. Therefore, other 
mechanisms for producing magnetic fields in white dwarfs have been recently
proposed. 

\citet{tou2008} proposed a binary origin where the magnetic field is formed
via a dynamo created during a common envelope (CE) phase. In systems that 
merge during the CE phase, single magnetic white dwarfs are created, but 
failed mergers would result in binary systems with a secondary nearly filling 
its Roche lobe. Following up on this theory, \citet{pot2010} and \citet{wic2014} 
showed that a magnetic field can be generated by a dynamo created by 
differential rotation within the CE, with the strongest fields
being created if the merged objects are differentially rotating near break-up.
A variation on the merger scenario was proposed by \citet{nor2011} who proposed 
that during a CE phase a low-mass star will be tidally 
disrupted by its proto-white dwarf companion forming an accretion disk. This would
generate a dynamo in the disk which is then transferred to the degenerate core via
accretion.
Magnetic fields can also be produced by the merger of two white dwarfs. \citet{gar2012}
have shown that the merger of two white dwarfs can generate a hot, convective and 
differentially rotating corona producing a dynamo and the resulting magnetic
field.  

\citet{ise2017} proposed that magnetic fields with $B \la 0.1$~MG may be
produced by phase separation during the onset of crystallization in the white 
dwarf core. They show that as white dwarfs begin to crystallize at sufficiently
low temperatures ($\sim 8000$~K), phase separation of the main elements (in 
most cases O and C) occurs leading to an unstable, convective liquid mantle on top of
a solid core. This produces a dynamo allowing the creation of a magnetic field.

\citet{bri2015} conducted a population synthesis of binary systems to investigate
which type of system could result in a magnetic white dwarf. They found that the
contribution from the double degenerate merger scenario is much smaller than 
the contribution from the CE merger.
Both merger scenarios are able to explain the higher than average mass of magnetic
white dwarfs. Once the field is established, the predicted magnetic field strengths
should remain throughout the white dwarf life-time since the magnetic flux is not
expected to decay significantly once the magnetic field is frozen into the 
white dwarf.

Evidence for the binary origin of magnetic fields in white dwarfs can be found 
in a few double degenerate systems. The fast rotating magnetic white dwarf 
EUVE~J0317-855 is in a common proper motion (CPM) binary with LP9802 \citep{kul2010} and has 
$B \approx 450$ MG magnetic spot with an underlying lower field of
$B \approx 185$ MG \citep{fer1997,ven2003}. The cooling age of 
EUVE~J0317-855 which is also the more massive white dwarf in the system
is much shorter than that of its CPM
companion LP9802 and therefore EUVE~J0317-855 is the result of a merger. The 
CPM binary PG1258+593 plus SDSS~J1300+5904 have similar 
masses, however SDSS~J1300+5904 is much cooler \citep{gir2010} resulting in an 
age discrepancy and therefore implying that PG1258+593 is the product of a 
merger.

Only a few magnetic plus non-magnetic double degenerate systems are known
\citep[for a list see][]{kaw2017}. In some cases, the magnetic white dwarf is
hotter and hence younger than its non-magnetic companion, despite being the more
massive component. Failed merger may also deliver double degenerate systems with a 
magnetic component. A magnetic field can be created during the
CE without the stars merging. This is the case of the close double
degenerate system NLTT~12758 which contains a magnetic white dwarf. The field
in the magnetic white dwarf was probably not formed during the merger of two 
stars, but would have formed in the second CE phase where the 
differential rotation was greater \citep{kaw2017}.

\section{Summary}

The growing sample of known magnetic white dwarfs is revealing a diversity in 
the properties of magnetic fields. The incidence of magnetism appears to vary
between various classes of white dwarfs. This suggests that magnetic fields in 
white dwarfs are created by several discernible processes.   

\acknowledgements
A.K. thanks S. Vennes and L. Ferrario for stimulating
discussions. This work was supported by the Czech Science Foundation (15-15943S).

\bibliography{kawka}

\begin{thebibliography}{44}
\expandafter\ifx\csname natexlab\endcsname\relax\def\natexlab#1{#1}\fi

%ADS_ID bri2015
\bibitem[{{Briggs} {et~al.}(2015){Briggs}, {Ferrario}, {Tout},
  {Wickramasinghe}, \& {Hurley}}]{bri2015}
{Briggs}, G.~P., {Ferrario}, L., {Tout}, C.~A., {Wickramasinghe}, D.~T., \&
  {Hurley}, J.~R. 2015, {\it \mnras}, {\bf 447}, 1713

%ADS_ID bri2005
\bibitem[{{Brinkworth} {et~al.}(2005){Brinkworth}, {Marsh}, {Morales-Rueda},
  {Maxted}, {Burleigh}, \& {Good}}]{bri2005}
{Brinkworth}, C.~S., {Marsh}, T.~R., {Morales-Rueda}, L., {et~al.} 2005, {\it
  \mnras}, {\bf 357}, 333

%ADS_ID bur1999
\bibitem[{{Burleigh} {et~al.}(1999){Burleigh}, {Jordan}, \&
  {Schweizer}}]{bur1999}
{Burleigh}, M.~R., {Jordan}, S., \& {Schweizer}, W. 1999, {\it \apjl}, {\bf
  510}, L37

%ADS_ID duf2005
\bibitem[{{Dufour} {et~al.}(2005){Dufour}, {Bergeron}, \& {Fontaine}}]{duf2005}
{Dufour}, P., {Bergeron}, P., \& {Fontaine}, G. 2005, {\it \apj}, {\bf 627},
  404

%ADS_ID duf2008
\bibitem[{{Dufour} {et~al.}(2008){Dufour}, {Fontaine}, {Liebert}, {Schmidt}, \&
  {Behara}}]{duf2008}
{Dufour}, P., {Fontaine}, G., {Liebert}, J., {Schmidt}, G.~D., \& {Behara}, N.
  2008, {\it \apj}, {\bf 683}, 978

%ADS_ID duf2013
\bibitem[{{Dufour} {et~al.}(2013){Dufour}, {Vornanen}, {Bergeron}, \&
  {Fontaine}}]{duf2013}
{Dufour}, P., {Vornanen}, T., {Bergeron}, P., \& {Fontaine}, A., B. 2013, in
  Astronomical Society of the Pacific Conference Series, Vol. 469, 18th
  European White Dwarf Workshop., 167

%ADS_ID dun2015
\bibitem[{{Dunlap} \& {Clemens}(2015)}]{dun2015}
{Dunlap}, B.~H. \& {Clemens}, J.~C. 2015, in Astronomical Society of the
  Pacific Conference Series, Vol. 493, 19th European Workshop on White Dwarfs,
  ed. P.~{Dufour}, P.~{Bergeron}, \& G.~{Fontaine}, 547

%ADS_ID far2011
\bibitem[{{Farihi} {et~al.}(2011){Farihi}, {Dufour}, {Napiwotzki}, \&
  {Koester}}]{far2011}
{Farihi}, J., {Dufour}, P., {Napiwotzki}, R., \& {Koester}, D. 2011, {\it
  \mnras}, {\bf 413}, 2559

%ADS_ID fer1997
\bibitem[{{Ferrario} {et~al.}(1997){Ferrario}, {Vennes}, {Wickramasinghe},
  {Bailey}, \& {Christian}}]{fer1997}
{Ferrario}, L., {Vennes}, S., {Wickramasinghe}, D.~T., {Bailey}, J.~A., \&
  {Christian}, D.~J. 1997, {\it \mnras}, {\bf 292}, 205

%ADS_ID gar2012
\bibitem[{{Garc{\'{\i}}a-Berro} {et~al.}(2012){Garc{\'{\i}}a-Berro},
  {Lor{\'e}n-Aguilar}, {Aznar-Sigu{\'a}n}, {Torres}, {Camacho}, {Althaus},
  {C{\'o}rsico}, {K{\"u}lebi}, \& {Isern}}]{gar2012}
{Garc{\'{\i}}a-Berro}, E., {Lor{\'e}n-Aguilar}, P., {Aznar-Sigu{\'a}n}, G.,
  {et~al.} 2012, {\it \apj}, {\bf 749}, 25

%ADS_ID gen2017
\bibitem[{{Gentile Fusillo} {et~al.}(2018){Gentile Fusillo}, {Tremblay},
  {Jordan}, {G{\"a}nsicke}, {Kalirai}, \& {Cummings}}]{gen2017}
{Gentile Fusillo}, N.~P., {Tremblay}, P.-E., {Jordan}, S., {et~al.} 2018, {\it
  \mnras}, {\bf 473}, 3693

%ADS_ID gir2010
\bibitem[{{Girven} {et~al.}(2010){Girven}, {G{\"a}nsicke}, {K{\"u}lebi},
  {Steeghs}, {Jordan}, {Marsh}, \& {Koester}}]{gir2010}
{Girven}, J., {G{\"a}nsicke}, B.~T., {K{\"u}lebi}, B., {et~al.} 2010, {\it
  \mnras}, {\bf 404}, 159

%ADS_ID hol2015
\bibitem[{{Hollands} {et~al.}(2015){Hollands}, {G{\"a}nsicke}, \&
  {Koester}}]{hol2015}
{Hollands}, M.~A., {G{\"a}nsicke}, B.~T., \& {Koester}, D. 2015, {\it \mnras},
  {\bf 450}, 681

%ADS_ID ise2017
\bibitem[{{Isern} {et~al.}(2017){Isern}, {Garc{\'{\i}}a-Berro}, {K{\"u}lebi},
  \& {Lor{\'e}n-Aguilar}}]{ise2017}
{Isern}, J., {Garc{\'{\i}}a-Berro}, E., {K{\"u}lebi}, B., \&
  {Lor{\'e}n-Aguilar}, P. 2017, {\it \apjl}, {\bf 836}, L28

%ADS_ID kaw2017
\bibitem[{{Kawka} {et~al.}(2017){Kawka}, {Briggs}, {Vennes}, {Ferrario},
  {Paunzen}, \& {Wickramasinghe}}]{kaw2017}
{Kawka}, A., {Briggs}, G.~P., {Vennes}, S., {et~al.} 2017, {\it \mnras}, {\bf
  466}, 1127

%ADS_ID kaw2004
\bibitem[{{Kawka} \& {Vennes}(2004)}]{kaw2004}
{Kawka}, A. \& {Vennes}, S. 2004, in IAU Symposium, Vol. 224, The A-Star
  Puzzle, ed. J.~{Zverko}, J.~{Ziznovsky}, S.~J. {Adelman}, \& W.~W. {Weiss},
  879--885

%ADS_ID kaw2006
\bibitem[{{Kawka} \& {Vennes}(2006)}]{kaw2006}
{Kawka}, A. \& {Vennes}, S. 2006, {\it \apj}, {\bf 643}, 402

%ADS_ID kaw2011
\bibitem[{{Kawka} \& {Vennes}(2011)}]{kaw2011}
{Kawka}, A. \& {Vennes}, S. 2011, {\it \aaa}, {\bf 532}, A7

%ADS_ID kaw2014
\bibitem[{{Kawka} \& {Vennes}(2014)}]{kaw2014}
{Kawka}, A. \& {Vennes}, S. 2014, {\it \mnras}, {\bf 439}, L90

%ADS_ID kaw2007
\bibitem[{{Kawka} {et~al.}(2007){Kawka}, {Vennes}, {Schmidt}, {Wickramasinghe},
  \& {Koch}}]{kaw2007}
{Kawka}, A., {Vennes}, S., {Schmidt}, G.~D., {Wickramasinghe}, D.~T., \&
  {Koch}, R. 2007, {\it \apj}, {\bf 654}, 499

%ADS_ID kep2013
\bibitem[{{Kepler} {et~al.}(2013){Kepler}, {Pelisoli}, {Jordan}, {Kleinman},
  {Koester}, {K{\"u}lebi}, {Pe{\c c}anha}, {Castanheira}, {Nitta}, {Costa},
  {Winget}, {Kanaan}, \& {Fraga}}]{kep2013}
{Kepler}, S.~O., {Pelisoli}, I., {Jordan}, S., {et~al.} 2013, {\it \mnras},
  {\bf 429}, 2934

%ADS_ID koe2009
\bibitem[{{Koester} {et~al.}(2009){Koester}, {Voss}, {Napiwotzki},
  {Christlieb}, {Homeier}, {Lisker}, {Reimers}, \& {Heber}}]{koe2009}
{Koester}, D., {Voss}, B., {Napiwotzki}, R., {et~al.} 2009, {\it \aaa}, {\bf
  505}, 441

%ADS_ID kul2010
\bibitem[{{K{\"u}lebi} {et~al.}(2010){K{\"u}lebi}, {Jordan}, {Nelan},
  {Bastian}, \& {Altmann}}]{kul2010}
{K{\"u}lebi}, B., {Jordan}, S., {Nelan}, E., {Bastian}, U., \& {Altmann}, M.
  2010, {\it \aaa}, {\bf 524}, A36

%ADS_ID lan2017
\bibitem[{{Landstreet} {et~al.}(2017){Landstreet}, {Bagnulo}, {Valyavin}, \&
  {Valeev}}]{lan2017}
{Landstreet}, J.~D., {Bagnulo}, S., {Valyavin}, G., \& {Valeev}, A.~F. 2017,
  {\it \aaa}, {\bf 607}, A92

%ADS_ID lan2012
\bibitem[{{Landstreet} {et~al.}(2012){Landstreet}, {Bagnulo}, {Valyavin},
  {Fossati}, {Jordan}, {Monin}, \& {Wade}}]{lan2012}
{Landstreet}, J.~D., {Bagnulo}, S., {Valyavin}, G.~G., {et~al.} 2012, {\it
  \aaa}, {\bf 545}, A30

%ADS_ID law2013
\bibitem[{{Lawrie} {et~al.}(2013){Lawrie}, {Burleigh}, {Dufour}, \&
  {Hodgkin}}]{law2013}
{Lawrie}, K.~A., {Burleigh}, M.~R., {Dufour}, P., \& {Hodgkin}, S.~T. 2013,
  {\it \mnras}, {\bf 433}, 1599

%ADS_ID lie2015
\bibitem[{{Liebert} {et~al.}(2015){Liebert}, {Ferrario}, {Wickramasinghe}, \&
  {Smith}}]{lie2015}
{Liebert}, J., {Ferrario}, L., {Wickramasinghe}, D.~T., \& {Smith}, P.~S. 2015,
  {\it \apj}, {\bf 804}, 93

%ADS_ID max2000
\bibitem[{{Maxted} {et~al.}(2000){Maxted}, {Ferrario}, {Marsh}, \&
  {Wickramasinghe}}]{max2000}
{Maxted}, P.~F.~L., {Ferrario}, L., {Marsh}, T.~R., \& {Wickramasinghe}, D.~T.
  2000, {\it \mnras}, {\bf 315}, L41

%ADS_ID nor2011
\bibitem[{{Nordhaus} {et~al.}(2011){Nordhaus}, {Wellons}, {Spiegel}, {Metzger},
  \& {Blackman}}]{nor2011}
{Nordhaus}, J., {Wellons}, S., {Spiegel}, D.~S., {Metzger}, B.~D., \&
  {Blackman}, E.~G. 2011, {\it Proceedings of the National Academy of Science},
  {\bf 108}, 3135

%ADS_ID pot2010
\bibitem[{{Potter} \& {Tout}(2010)}]{pot2010}
{Potter}, A.~T. \& {Tout}, C.~A. 2010, {\it \mnras}, {\bf 402}, 1072

%ADS_ID put1997
\bibitem[{{Putney}(1997)}]{put1997}
{Putney}, A. 1997, {\it \apjss}, {\bf 112}, 527

%ADS_ID sch1995
\bibitem[{{Schmidt} \& {Smith}(1995)}]{sch1995}
{Schmidt}, G.~D. \& {Smith}, P.~S. 1995, {\it \apj}, {\bf 448}, 305

%ADS_ID tou2008
\bibitem[{{Tout} {et~al.}(2008){Tout}, {Wickramasinghe}, {Liebert}, {Ferrario},
  \& {Pringle}}]{tou2008}
{Tout}, C.~A., {Wickramasinghe}, D.~T., {Liebert}, J., {Ferrario}, L., \&
  {Pringle}, J.~E. 2008, {\it \mnras}, {\bf 387}, 897

%ADS_ID tre2015
\bibitem[{{Tremblay} {et~al.}(2015){Tremblay}, {Fontaine}, {Freytag},
  {Steiner}, {Ludwig}, {Steffen}, {Wedemeyer}, \& {Brassard}}]{tre2015}
{Tremblay}, P.-E., {Fontaine}, G., {Freytag}, B., {et~al.} 2015, {\it \apj},
  {\bf 812}, 19

%ADS_ID val2014
\bibitem[{{Valyavin} {et~al.}(2014){Valyavin}, {Shulyak}, {Wade}, {Antonyuk},
  {Zharikov}, {Galazutdinov}, {Plachinda}, {Bagnulo}, {Fox Machado}, {Alvarez},
  {Clark}, {Lopez}, {Hiriart}, {Han}, {Jeon}, {Zurita}, {Mujica}, {Burlakova},
  {Szeifert}, \& {Burenkov}}]{val2014}
{Valyavin}, G., {Shulyak}, D., {Wade}, G.~A., {et~al.} 2014, {\it \nat}, {\bf
  515}, 88

%ADS_ID val2008
\bibitem[{{Valyavin} {et~al.}(2008){Valyavin}, {Wade}, {Bagnulo}, {Szeifert},
  {Landstreet}, {Han}, \& {Burenkov}}]{val2008}
{Valyavin}, G., {Wade}, G.~A., {Bagnulo}, S., {et~al.} 2008, {\it \apj}, {\bf
  683}, 466

%ADS_ID ven2003
\bibitem[{{Vennes} {et~al.}(2003){Vennes}, {Schmidt}, {Ferrario}, {Christian},
  {Wickramasinghe}, \& {Kawka}}]{ven2003}
{Vennes}, S., {Schmidt}, G.~D., {Ferrario}, L., {et~al.} 2003, {\it \apj}, {\bf
  593}, 1040

%ADS_ID vor2013
\bibitem[{{Vornanen} {et~al.}(2013){Vornanen}, {Berdyugina}, \&
  {Berdyugin}}]{vor2013}
{Vornanen}, T., {Berdyugina}, S.~V., \& {Berdyugin}, A. 2013, {\it \aaa}, {\bf
  557}, A38

%ADS_ID wic2000
\bibitem[{{Wickramasinghe} \& {Ferrario}(2000)}]{wic2000}
{Wickramasinghe}, D.~T. \& {Ferrario}, L. 2000, {\it \pasp}, {\bf 112}, 873

%ADS_ID wic2005
\bibitem[{{Wickramasinghe} \& {Ferrario}(2005)}]{wic2005}
{Wickramasinghe}, D.~T. \& {Ferrario}, L. 2005, {\it \mnras}, {\bf 356}, 1576

%ADS_ID wic2014
\bibitem[{{Wickramasinghe} {et~al.}(2014){Wickramasinghe}, {Tout}, \&
  {Ferrario}}]{wic2014}
{Wickramasinghe}, D.~T., {Tout}, C.~A., \& {Ferrario}, L. 2014, {\it \mnras},
  {\bf 437}, 675

%ADS_ID zuc2011
\bibitem[{{Zuckerman} {et~al.}(2011){Zuckerman}, {Koester}, {Dufour}, {Melis},
  {Klein}, \& {Jura}}]{zuc2011}
{Zuckerman}, B., {Koester}, D., {Dufour}, P., {et~al.} 2011, {\it \apj}, {\bf
  739}, 101

%ADS_ID zuc2003
\bibitem[{{Zuckerman} {et~al.}(2003){Zuckerman}, {Koester}, {Reid}, \&
  {H{\"u}nsch}}]{zuc2003}
{Zuckerman}, B., {Koester}, D., {Reid}, I.~N., \& {H{\"u}nsch}, M. 2003, {\it
  \apj}, {\bf 596}, 477

%ADS_ID zuc2010
\bibitem[{{Zuckerman} {et~al.}(2010){Zuckerman}, {Melis}, {Klein}, {Koester},
  \& {Jura}}]{zuc2010}
{Zuckerman}, B., {Melis}, C., {Klein}, B., {Koester}, D., \& {Jura}, M. 2010,
  {\it \apj}, {\bf 722}, 725

\end{thebibliography}
\end{document}